# Partially Informed Elections:
## Analyzing the Impact of Forced Ballot Truncation on Bucklin, Coombs, Plurality with Runoff, and Schulze

Jonah Stein

Independent Researcher

jonah.s.stein@gmail.com

**Abstract.** Elections employ various voting systems to determine winners based on voters' preferences. However, many recent ranked-choice elections have forced voters to truncate their ballots by only ranking a subset of the candidates. This study analyzes how forced ballot truncation affects the Bucklin, Coombs, plurality with runoff, and Schulze voting systems' abilities to output their true winning sets. Using computer simulations, thousands of preference profiles were generated with the Mallows model using different numbers of candidates, voters, and dispersion values. The true winning set was determined for each system using complete preferences, then compared to winning sets derived from repeatedly truncated preferences within the same preference profile. Results show that plurality with runoff was the most resistant to forced truncation, followed by Schulze, Bucklin, and Coombs. Additionally, elections with fewer voters and higher dispersion values were found to decrease the probability of selecting the true winning set across all systems. The findings provide insights into how forced truncation impacts voting systems, aiding election designers in their work.

## 1. Intro:

Elections are everywhere in the modern world most obviously, perhaps, in the political sphere, where citizens will vote for their representatives and country leaders. However, elections take place in many other environments, such as within businesses, charities, and even reality tv. All of these elections select their winners using voting systems, methods used to "select a winner (or winners) from a set of alternatives taking into account everyone's opinion"(Pacuit 2019). The majority of these elections determine their winners using plurality voting, a voting system that allows voters to vote for one candidate and outputs the candidate who received the most votes (Pacuit 2019). Although this voting system may seem logical, it has issues that are leading people to look for other voting systems. One of plurality's main issues is its vulnerability to "spoiler candidates" -- candidates with no realistic chance of winning the election but whose presence in the election may affect the outcome of it(Shugart 2004). Critics of plurality also argue that, due to the one-preference nature of plurality, voters will avoid casting ballots for smaller candidates, despite them being their first preference. Instead, they will vote for candidates who have a better chance of winning the election to prevent their vote from being "wasted" (Smith 2006). These two issues produced by the plurality voting system can produce election results that are not truly representative of voters' preferences, prompting discussion about new voting methods to fill these gaps.

To combat some of plurality's issues, social choice theorists have fabricated new methods that utilize varied processes to determine the winners of elections. Within America's political sphere, the most widely adopted alternative voting method has been ranked choice voting(referred to as RCV), being used by 62 jurisdictions within the United States as of December 2022(Vasilogambros 2022). RCV employs preferential voting, which allows voters to rank candidates in order of their preference. Using these preferential ballots, RCV eliminates the candidate with the least amount of 1st preference votes. Then, the ballots that had ranked the eliminated candidate highest get transferred to their next most preferred candidate, and the cycle repeats until one candidate obtains a majority of the votes(John 2017). Proponents of RCV argue that it produces more representative outcomes because voters can better express their preferences(Campaign Legal Center). Alternatively, opponents claim that in some cases, voters can provide less input into the outcome due to their ballots becoming exhausted(Congressional Digest 2022). The criticism and praise received by RCV highlight the statute that a perfect voting system does not exist.

## 2. Literature Review:

### 2.1. Voting System's Criteria:

As mentioned above, choosing a perfect voting system is not possible, an idea by economist and Nobel Laureate Kenneth Arrow in what is now termed Arrow's Impossibility Theorem, which states that no voting system can satisfy every criterion upon which it is judged(Radcliffe 2018). Because of this theorem, when deciding which voting method to use, election designers must consider and weigh various criteria against each other.

One of the most common and highly weighted voting criteria is the Condorcet criterion. This criterion states that, when possible, the winner of an election should be the candidate that would defeat all other candidates in a head-to-head election(Bag 2009).

Another common voting criterion is the monotonicity criterion. A voting method satisfies this criterion if an additional vote ranking a candidate higher than another does not hurt the higher-ranked candidate's standings and vice versa(Radcliffe 2018). For example, if an additional vote ranks candidate A over candidate B, candidate A's overall standing in the election should not be hurt.

One more example of a voting criterion is the majority criterion. This criterion is satisfied if a voting method always outputs the candidate ranked first by more than 50% of the voters(Rothe 2015).

There is a slew of other criteria used to judge voting systems on other characteristics, including whether they are affected by "spoiler" candidates or whether the Condorcet loser, the candidate ranked last more often than any other candidate in a head-to-head, can be elected. Additionally, due to the prevalence of incomplete ballots, social choice theorists should consider a voting system's susceptibility to truncated ballots when evaluating it against others.

### 2.2. Why Voters Truncate Ballots

There are many reasons that explain why elections contain truncated ballots. A truncated ballot is a ballot in a preferential voting system that does not rank every candidate in an election(Kilgour et al. 2019). One explanation social choice theorists point to is that voters can truncate ballots to manipulate election results(Saari and Van Newenhizen 1988). This is referred to as the truncation paradox and has been studied by many social choice theorists to determine which voting systems encounter it the most and are most vulnerable to it.(Kamwa 2022, Kamwa and Moyouwou 2021, Nurmi 1999, Plassmann and Tideman 2013). Partially due to this paradox and to avoid exhausted ballots, or ballots that run out of ranked candidates and are therefore not counted, some elections, such as Australia's congressional elections(Owen 2018), do not count incomplete ballots.

Although ignoring incomplete ballots may abolish the chance of a truncation paradox, many scholars argue that truncated ballots should count in elections. Ayadi et al. show that voters, when able, prefer to provide ballots with partial rankings(Ayadi et al. 2019). Supporting this stance against the compulsory ranking of candidates, Yuval Filmus, an associate professor at Technion, and Joel Oren, who received a Ph.D. in Computer Science from the University of Toronto, state that "In a system with a vast collection of candidates to choose from, obtaining a [voter's] complete ranking of the candidates is often ill-advised, and even infeasible, due to the resulting communication and cognitive overhead."(Filmus and Oren 2014).

Additionally, many elections force truncate ballots by only allowing a certain number of candidates to be ranked. For example, in the Democratic mayoral primaries in New York City, voters can only rank five candidates(NYC Civic engagement Commission). Similarly, in preferential elections in Alameda County, California, voters can only rank three candidates (Alameda County Government). The reasons for this forced truncation are fuzzy, as Kilgour et al. write that it "seems arbitrary and perhaps simply a matter of convenience for officials running the election"(Kilgour et al. 2019). Despite the reasoning behind forced ballot truncation and ballot truncation in general, it is an obstacle that voting systems will have to hurdle, so it is important to study their resistance to it.

## 2.3. Related Work

Current research concerning the effects of truncated ballots on election outcomes mainly focuses on RCV elections. For instance, D. Marc Kilgour, a professor of Mathematics at Wilfrid Laurier University, et al. simulated RCV election results with complete ballots and then uniformly truncated the ballots of those results to determine how the Condorcet efficiency(how often the Condorcet candidate was outputted as the winner) and winners were affected by forced ballot truncation. They found that the Condorcet efficiency of RCV was low without truncation, but was even worse with the force-truncated ballots. The researchers also found that the level of forced truncation heavily impacted the election results(Kilgour et al. 2019). Consistent with this previous study, Kiran Tomlinson, a computer science Ph.D. student at Cornell University, et al. found that the chance that the winner of an RCV election with force-truncated ballots was the same winner as that of an identical election with complete ballots decreased as ballot length decreased. Interestingly, the researchers observed that ballots filled with at least half of the candidates almost always produced the same results as full RCV ballots(Tomlinson et al. 2022). Both studies exhibit the significant effect forced ballot truncation has on RCV elections.

In a broader scope, Manel Ayadi, a postdoctoral researcher focused on voting systems, explored the impacts of truncated ballots on election outcomes of various voting systems including Copeland, Harmonic, Borda, Maximin, Ranked Pairs, and single transferable vote(equivalent to RCV). The study found that with all of the voting systems, the original result of the election with complete ballots was able to be obtained with a relatively high level of truncation when elections were simulated with low dispersion values. Ayadi also found that Harmonic voting was able to produce the best prediction of the winner with force-truncated ballots(Ayadi 2019). One explanation for the difference in findings from Ayadi and the studies by Tomlinson et al.(2022) and Kilgour et al.(2019) is the method used to simulate the elections. Ayadi simulated elections with voter bases with closer preferences to each other, whereas Tomlinson et al. and Kilgour et al. generated voter bases with random preferences.

## 2.4. The Gap

The previous research exploring the effects of forced ballot truncation on voting systems primarily focused on RCV's reaction to it. Ayadi researched the effects of force-truncated ballots on voting systems other than RCV(2019), but there are still many voting systems whose reactions to force-truncated ballots are unknown. This study intends to address this gap by answering the question: "How does forced ballot truncation affect the ability of the Bucklin, Coombs, plurality with runoff, and Schulze voting methods to output their true winners?" As discussed above, forced ballot truncation is a common occurrence in current preferential elections, so measuring a voting system's response to force-truncated ballots can add another data point for voting systems to be judged on. This research question can also be seen as an extension of Baumeister et al.'s work, as they suggest that future research explore "how likely it is that the result is determined (i.e., that there exists a *necessary* winner) when all voters have specified ballots of a given size?"(Baumeister et al. 2012). By extending this research, election designers will be more equipped to decide which voting system to use when using force-truncated ballots.

## 3. Methodology

### 3.1. Research Method

To measure the effects of forced ballot truncation on various voting systems, this study employs a computer simulation, a method used throughout the field of study(Kilgour et al. 2019, Ayadi et al. 2019, Tomlinson et al. 2022). The use of a computer simulation allows for thousands of elections to be quickly simulated and analyzed, which enables the study to reach more substantial conclusions from a larger sample set. Other methods of research including analyzing the voting systems with empirical data or creating new elections with real participants would produce fewer data points and yield smaller sample sizes, resulting in less significant findings.

### 3.2. Voting Systems

Four voting systems were considered in this study: Bucklin, Coombs, plurality with runoff, and Schulze. These voting systems were chosen due to their respectively unique determination methods and the absence of research regarding their response to force-truncated ballots. Both plurality with runoff and Schulze are designed to intake truncated ballots, whereas Bucklin and Coombs were both adapted to work with them. For the sake of this study, the "winning set" refers to all of the winners of a certain election.

Plurality with runoff first determines the two candidates(or more if there are ties) with the most first-place rankings. If a candidate is ranked first by more than 50% of the voters, then that candidate is declared the winner. If there is no candidate with a strict majority of first-place votes, then there is a runoff(using the remaining preferences) between the remaining candidates, and the winner(s) of the runoff is deemed the winning set of the election(Pacuit 2019). The Schulze method determines its winning set based on the margin of preference between each candidate (Schulze 2010).

Bucklin and Coombs do not work with truncated ballots, so they were adapted to handle truncated ballots while retaining their respective general determination methods. These adaptations were made after consulting with Eric Pacuit, a logic professor at the University of Maryland who has devoted much of his research to voting systems.

Bucklin usually works by continually counting the level of preferences on ballots until one candidate has reached at least a strict majority of voters(Smith 2006). For example, if after all the first preferences are tallied, a candidate does not have a strict majority(>50%) of the votes, then the second preference votes are counted. Because truncated ballots can result in a strict majority never being found, Bucklin was adapted so that if all levels of the ballot were counted, and a strict majority had not been achieved by any of the candidates, then the candidate(s) with the highest score was outputted as the winning set.

Coombs usually works by eliminating the candidate with the most last-place votes until there is a candidate that has a strict majority(>50%) of first-place votes(Grofman and Feld 2004). Because force-truncated ballots can prevent Coombs from ever finding a candidate that has a strict majority of first-place votes, the system was adapted so that the winning set is comprised of the candidates that are eliminated last(if a candidate does not obtain a strict majority of first preference votes first).

### 3.3 Generating Profiles

This study uses the mallows $\phi$ probability model to generate preference profiles(referred to as profiles throughout the rest of this paper) for a population of voters. A preference profile is a collection of the voters' preferences(or rankings) in an election. The mallows $\phi$ probability model first generates a random reference ranking of the candidates in an election(i.e. Candidate A > Candidate B > Candidate C). Next, the model generates other rankings based on a dispersion parameter $\phi$(ranges from 0-1) and the reference ranking. The value of the dispersion parameter determines the similarity of preferences of a certain population of voters. For example, a profile generated with $\phi$=0 will result in a profile in which all the voters have identical preferences and are thus assigned the reference ranking. On the other end, a profile generated with $\phi$=1 will result in a random distribution of preferences. In other words, "with small values of $\phi$, the mass [of voters] is concentrated around [the reference ranking]," while larger values of $\phi$ result in profiles with more distributed rankings(Ayadi 2019). These profiles are then used as inputs for the voting systems above, which each output their winning set.

Using the mallows $\phi$ probability model, profiles were generated with varying numbers of voters, candidates, and dispersion values. Aligning with other studies, profiles were generated with 4,5,6, and 7 candidates. These values were chosen because Kilgour et al. studied the same topic by simulating elections with 4,5, and 6 candidates(Kilgour et al. 2019). Additionally, Ayadi et al. used simulations containing 7 candidates(Ayadi et al. 2019). Furthermore, the above studies simulated their elections with voter bases containing 100-601 voters with 100 voter increments(ie. 100 voters, 200 voters,...). Thus, this study simulates elections with 100, 200, 300, 400, 500, and 600 voters. This study also simulates elections with 2000 voters, a value based on Manel Ayadi's study(Ayadi 2019), to determine the effect of forced ballot truncation on each voting system for large elections.

Profiles were also generated with dispersion values of 0.7, 0.8, 0.9 based on Ayadi et al.'s use of these values(Ayadi 2019). A dispersion value of 1 was also used in this study to simulate profiles with a random distribution of rankings, as used by both Kilgour et al. and Ayadi et al.(Kilgour et al. 2019, Ayadi et al. 2019).

### 3.4. Running the Simulation

The Preferential Voting Tools python library, developed by Eric Pacuit, a logic professor at the University of Maryland, and Wes Holiday, a philosophy professor at the University of California, Berkeley, was used to generate profiles using the mallows $\phi$ model(Pacuit and Holiday). The python library also included code for the Schulze(under the function name beat_path) and plurality with runoff voting systems. The adapted Bucklin and Coombs voting systems were re-programmed using methods embedded in the python library.

For each set of inputs(number of candidates, number of voters, and dispersion value), 1000 profiles were generated. This repetition value is based on Ayadi et al.'s study, in which they repeated each permutation of inputs 1000 times as well(Ayadi et al. 2019). This amount of repetition provides a large enough sample set to make substantiated claims while limiting the computational demand. The code used to run the simulations and adapt the voting methods can be found on GitHub (Anonymous Github 2023).

### 3.5. Measuring and Analyzing the Data

The metric used to determine the effect of forced ballot truncation on each voting system is based on Ayadi and Kilgour et al.'s studies in which they measured the rate at which voting systems select their "true" winning set at each level of truncation(Ayadi 2019, Kilgour et al. 2019). For each simulated profile, first, the "true" winning set for each voting system was determined by selecting the winner(s) of the election with the complete preferences. Next, subsequent elections were conducted using the same profile but with all of the ballots truncated by one candidate, then two, until only one candidate was left. At each level of truncation, each voting method's newly determined winning set was compared against its "true" winning set. If the two sets were equivalent, then the score for the voting system at the level of truncation at which it occurred was incremented by 1. This process was repeated 1000 times for each permutation of profile inputs.

To record and analyze the data, a graph for each set of parameters was created. This is consistent with the strategy used by Ayadi et al.(2019) as well as Kilgour et al.(2019). For each graph, the independent variable, the level of truncation, was plotted against the probability in which the "true" winner was selected for all four voting systems. These graphs were then analyzed to make conclusions regarding each of the voting system's resistance to ballot truncation. Furthermore, all of the data was recorded in a spreadsheet to be further analyzed(See **Appendix A**).

## 4. Results

After simulating thousands of preference profiles with differing values of candidates, voters, and dispersion, the data was recorded to a spreadsheet and organized to perform greater analysis.

Using data from the simulations, graphs were generated to plot the probability that the new winning set generated by each voting system at each truncation level was equivalent to the true winning set. The probabilities plotted on the vertical axis were calculated by summing the number of instances that the winning set under each level of truncation was equal to its true winning set divided by the number of simulation runs. For example, if 1000 simulations were run, and the winning set at a certain truncation level is equal to the true winning set 800 times, the probability would be 800/1000, or 0.8. These probabilities were plotted for all four voting systems for each level of truncation.

On each graph, the Coombs voting system is represented by the blue line, the Bucklin voting system by the red line, plurality with runoff by the yellow line, and Schulze by the green line. Truncation values for all of the graphs started at 2, as truncating a profile by one candidate does not change the outcome of the election, as the unranked candidate can be assumed to be ranked last.

### Figure 1

True Winner Set Chosen vs. Truncation Level

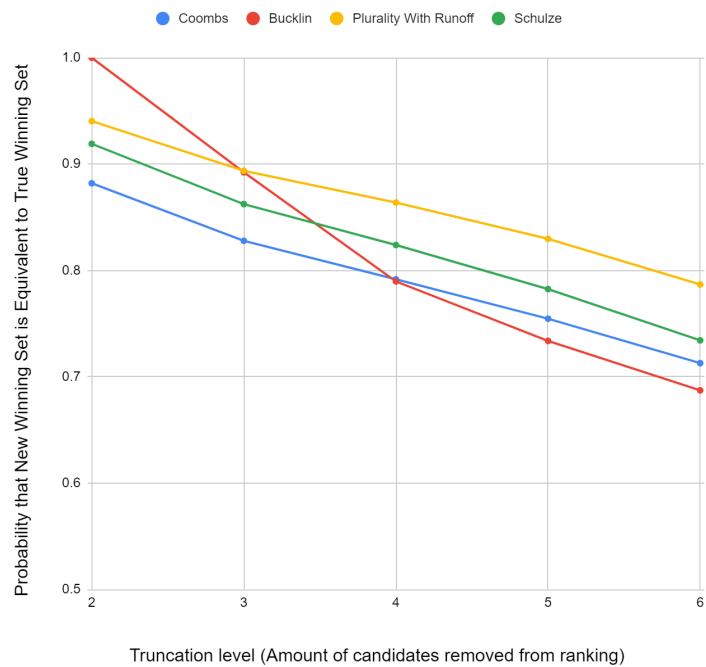

**Figure 1** represents data from all of the simulations, accounting for all permutations of candidates, numbers of voters, and dispersion values. The data were then organized by the number of voters, dispersion value, and amount of candidates to analyze the impacts of each of the variables. **Figure 2** compares the probabilities that each voting method's winning set under each level of truncation is equivalent to its true set between profiles with smaller and larger numbers of voters. **Figure 3** compares the same relationship between simulations with higher and lower dispersion values. Lastly, **Figure 4** compares the same relationship, but with simulations containing varying numbers of candidates.

### Figure 2

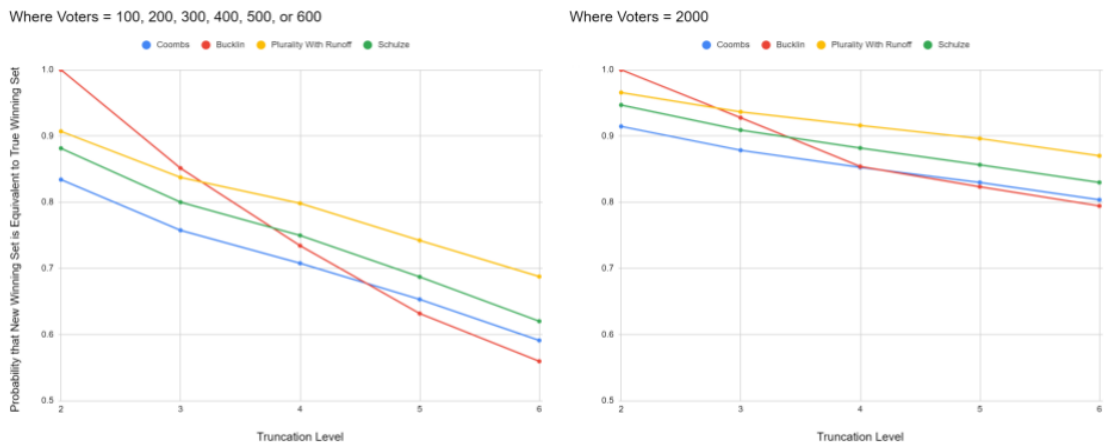

Compares the differences between the probabilities that the voting systems choose the true winner set under each level of truncation with a smaller amount of voters(left) versus a larger amount of voters(right)

### Figure 3

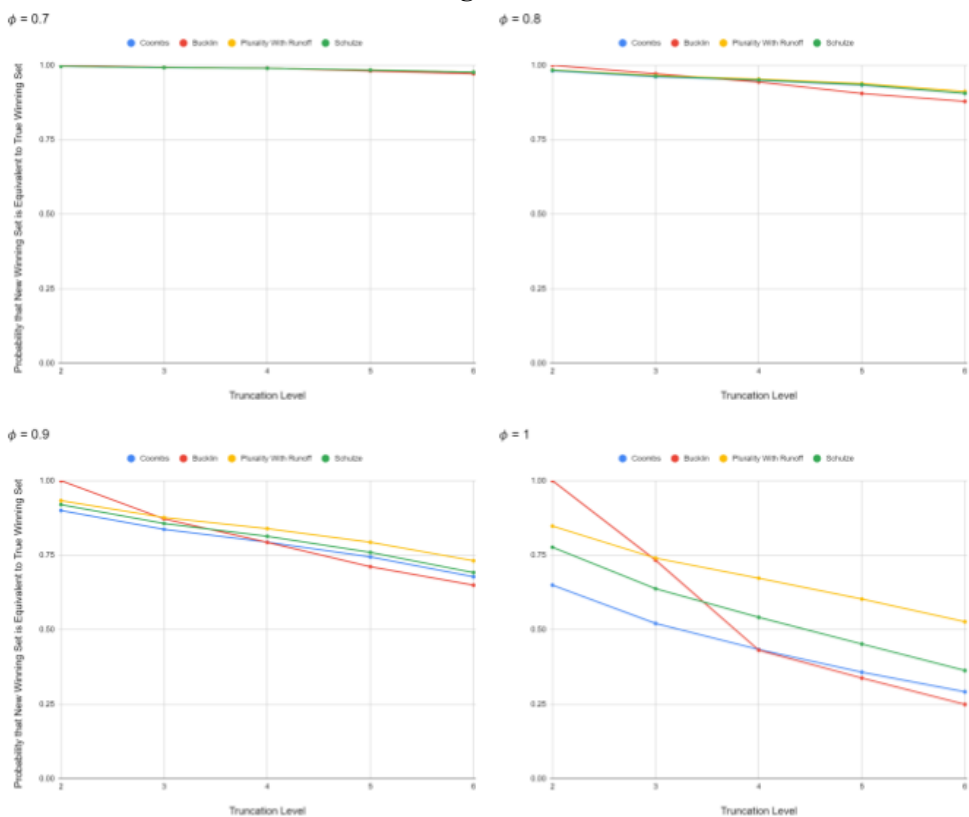

Compares the differences between the probabilities that the voting systems choose the true winner set under each level of truncation using data from preference profiles generated with different dispersion values:
$\phi = 0.7$(top left),  $\phi = 0.8$(top right),  $\phi = 0.9$(bottom left),  $\phi = 1$(bottom right)

**Figure 4**

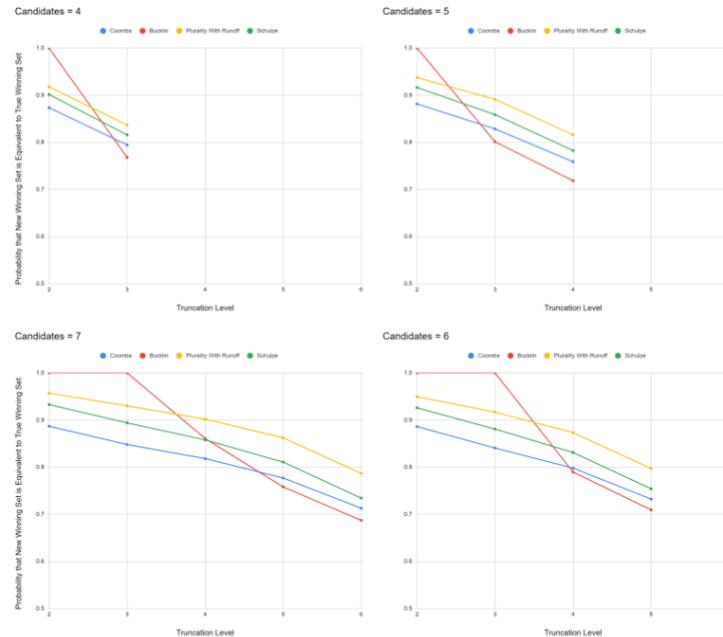

Compares the differences between the probabilities that the voting systems choose the true winner set under each level of truncation using data from preference profiles generated with different amounts of candidates:
4 candidates(top left), 5 candidates(top right), 6 candidates(bottom right), 7 candidates(bottom left)

## 5. Discussion and Analysis

In this section, two overarching topics will be analyzed: the factors affecting the results and the effect that forced ballot truncation had on each of the voting systems. The three variables altered throughout the simulation(number of voters, dispersion value, and the number of candidates) impacted the results in different and consistent ways that aid in understanding the effects of truncation within elections.

### 5.1. Voters

For every voting system, the number of voters in the election significantly contributed to the probability that each selected its true winning set. As shown in **Figure 2**, the voting systems had a smaller chance of electing the true winning set when the election contained fewer voters. For example, averaging the results from all levels of truncation and all voting methods, elections with 100, 200, 300, 400, 500, and 600 voters had a 0.75 chance of selecting the true winning set versus a 0.88 chance with 2000 voters(~17% increase). This correlation between the number of voters and the probability that the true winning set is selected is consistent with other studies by Bentert and Skowron(2020) and Ayadi(2019), in which larger sets of voters also contributed to lower probabilities in which the true winning set is chosen. This relationship is logical given that this study employed the mallows model to generate preference profiles. Because of this, as the voter base expands, it is more likely that more preferences will match, or at least be closer to, the reference ranking, thus leading the preferences to converge to the reference ranking.

Another interpretation of this pattern could justify the forced ballot truncation in the aforementioned elections in New York City Alameda County because their large voter bases would make the force-truncated ballots less likely to affect the outcomes. Although the results from this study support the above conclusion, more research is needed to validate this justification.

## 5.2. Dispersion

Another factor that affected the probability of the true winning set being chosen was the dispersion value inputted into the mallows model. **Figure 3** illustrates how larger dispersion values decreased the probability that the true winning set was chosen for every voting method. This relationship was also found by Ayadi(2019) when she measured the probability in which the true winner was chosen by various voting methods with varying dispersion values of 0.9 and 1.

More specifically, when profiles were simulated with dispersion values of 0.7, all the voting methods were able to select their true winning sets in >97% of the elections, even with maximum truncation(only 1 candidate on the ballot).

Dispersion values of 0.8 yielded similarly high results, with all the voting systems except for Bucklin being able to select their true winning sets >90% time(Bucklin chose the correct winner ~88% of the time with ballots truncated by 6).

When the dispersion values increased to 0.9 and 1, the probabilities that the voting methods select their true winners decreased dramatically to around an average of 69% for all voting systems at 0.9, and an average of 36% for all voting systems at 1.

## 5.3. Candidates

As the number of candidates increased, the probability of each voting system outputting its true winning set at shortened lengths decreased. However, it is worth noting that the number of candidates did not affect the results as much as the dispersion value. On average, each additional candidate decreased the chance that a voting method outputs its true winning set by ~2%. Similar results were found by Kilgour et al.(2019) in which they found that, for ranked choice voting, the probability that the winner chosen under each level of truncation differed from the true winner increased by a near-constant amount of ~5% for each additional candidate. The slight disparity between these two values can be attributed to two factors. Primarily, Kilgour et al.(2019) received those results using a random model to generate preference profiles(equivalent to a profile generated with the mallows model with $\phi = 1$), whereas the preference profiles in this study were simulated with less dispersion, which would decrease the effects of additional candidates, as preferences are more aligned to begin with.

## 5.4. Methods not designed for truncation

In their original forms, both Coombs and Bucklin require voters to rank all of the candidates on their ballots (Yonk et al. 2010), so they were adapted to handle truncated ballots. On average, the adapted versions of these voting systems performed worse than Schulze and plurality with runoff, the voting systems designed to handle incomplete ballots. Nonetheless, there were some notable findings for the two voting systems.

## 5.5 Coombs

Perhaps one of the biggest surprises to come from this study is Coomb's ability to output its true winning set under increasing levels of truncation. Due to Coomb's method of eliminating candidates with the most last-place votes, it was expected that the method would woefully underperform, especially compared to the rest of the voting systems. This reasoning is supported by Robert Loring, a long-time researcher of voting systems, who concluded that Coombs was one of the most sensitive voting methods to incomplete ballots(Loring). This study found that in the simulations, Coombs chose its true winner, on average, ~2.5% less than Bucklin, ~3% less than Schulze, and ~7% less than Plurality with Runoff. However, when ballots were truncated to at least their halfway point minus 1(for odd values of candidates), Coombs was able to output its true winner ~3% more than Bucklin.

**5.6 Bucklin**

Bucklin is another voting system that normally requires all candidates to be ranked, however, because of the method that it uses to determine its winner, it only needs just over half of the preferences to be tallied to find its true winner(ie. In an election with 7 candidates, tallying up to 4 preferences of every ballot will guarantee that the system's true winner be outputted). This study found that Bucklin needed even fewer preferences, finding that it would output its true winner whenever at least half of the candidates were ranked. Although it is technically possible for Bucklin to output a different winning set with only half the candidates ranked(for elections with even numbers of candidates), it did not occur over the 112,000 simulated elections in this study. Despite Bucklin's ability to choose its true winning set with lower levels of truncation, once the profiles were truncated below half of the candidates, Bucklin performed, on average, worse than every other voting system; Bucklin outputted its true winner ~2% less often than coombs, ~4% less often than Schulze, and ~8% less often than Plurality with Runoff.

**5.7 Plurality With Runoff**

Plurality with runoff performed the best among the 3 other voting systems considered in this study, with its rate of selecting its true winner being 86% compared to 79%, 82%, and 82% for Coombs, Bucklin, and Schulze, respectively. This result is reasonable given that within the first round of plurality with runoff, the same candidates move on to the next round. After that, the candidate pool would already be limited, giving the voting method a higher chance of selecting its true winning set. Despite being the most resistant to truncated ballots, plurality with runoff has downfalls that make it unfavorable such as failing to always elect the Condorcet winner(Wright and Riker 1989), being vulnerable to monotonicity paradoxes(instances in which a candidate being ranked higher by voters results in that candidate performing worse)(Lepelley 1996), and other paradoxes as well.

**5.8 Schulze**

Schulze was the only Condorcet-consistent voting system researched in this study, and it was the second-least resistant to truncated ballots, being able to select its winning set ~0.4% more often than Bucklin and ~3% more often than Coombs. Schulze's vulnerability to truncation is partially predetermined due to its Condorcet consistency. Because Condorcet winners are found using the margins between each candidate, when more candidates are excluded from rankings, there are fewer margins between candidates, which impacts the Condorcet winner. Therefore, although the results of this study do provide a sense of Schulze's ability to output its true winning set with truncated ballots, future research could address the differences in the abilities of other Condorcet-consistent voting systems to output their true winning sets when faced with force-truncated ballots.

## 6. Conclusion:

The purpose of this study was to determine the ability of the Bucklin, Coombs, plurality with Runoff, and Schulze voting methods' resistance to force-truncated ballots, by measuring each voting method's ability to output its true winning set at each level of truncation. After simulating 112,000 elections, this study ultimately determined that plurality with runoff was the least resistant voting system to forced ballot truncation, followed by Schulze, then Bucklin, and lastly, Coombs. The data and findings produced in this study provide a greater understanding of each of the voting methods and their reactions to forced ballot truncation; however, a few limitations have to be addressed.

First, when generating elections with the mallows $\phi$ model the resulting preference profiles were not stored. This closed off opportunities for deeper analysis, as the saved profiles could've provided insights into which specific voter population behaviors resulted in each voting system's relative performance. Although this limitation was partially mitigated by the use of varied dispersion values, the specific composition of each profile couldn't be determined. Future research could address this gap by rerunning this study's simulation, but storing each profile to save for deeper analysis. Additionally, future studies could use alternative algorithms to generate preference profiles such as impartial anonymous culture or spatial models.

Another potential limitation of this study was the metric used to determine a voting method's resistance to forced ballot truncation. In this study, each voting method's resistance to truncated ballots was determined by its ability to output its true winning set with different levels of truncation on the same preference profiles, a choice made in alignment with research by Ayadi, Bentert and Skowron, and Skowron et al.. Future research could measure the resistance of Bucklin, Coombs, plurality with runoff, Schulze, and other voting methods with other metrics such as the rate at which they choose their Condorcet winner(s), a metric employed by Kilgour et al.(2019), or their ability to avoid the Condorcet loser(s) at each level of truncation.

Lastly, this study simulated elections with force-truncated ballots while assuming that all of the ballots were completely filled out. This limits the scope of this study's conclusions because voters can be faced with force-truncated ballots and still choose to leave them incomplete. Future research could fill this gap by researching the effects that incompletely filled, force-truncated ballots have on different voting systems. Both Kilgour et al.(2019) and Ayadi et al.(2019) explore the effects of probabilistic truncation(varied levels of truncation based on empirical data) on ranked choice voting, however, this framework has not yet been combined with forced ballot truncation. Findings from this specific research can aid in even more implications, as it generates data from more realistic circumstances.

The limited conclusions reached from this research combined with both previous and future research on other voting methods' reactions to various types of ballot truncation can lead to more purposefully designed elections with better-understood voting systems. Election designers can leverage this study to determine the right voting system based on their voter population, distribution, and level of forced ballot truncation.

**Appendix A**

| Candidates | Voters | $\phi$ | Voting system | Probability that the true winning set is chosen at each ballot length(L) | | | | | | |
|---|---|---|---|---|---|---|---|---|---|---|
| | | | | L=1 | L=2 | L=3 | L=4 | L=5 | L=6 | L=7 |
| 4 | 100 | 0.70 | Bucklin | 0.87 | 1.00 | 1.00 | 1.00 | n/a | n/a | n/a |
| 4 | 100 | 0.70 | Coombs | 0.92 | 0.97 | 1.00 | 1.00 | n/a | n/a | n/a |
| 4 | 100 | 0.70 | Plurality With Runoff | 0.93 | 0.97 | 1.00 | 1.00 | n/a | n/a | n/a |
| 4 | 100 | 0.70 | Schulze | 0.93 | 0.97 | 1.00 | 1.00 | n/a | n/a | n/a |
| 4 | 100 | 0.80 | Bucklin | 0.75 | 1.00 | 1.00 | 1.00 | n/a | n/a | n/a |
| 4 | 100 | 0.80 | Coombs | 0.81 | 0.90 | 1.00 | 1.00 | n/a | n/a | n/a |
| 4 | 100 | 0.80 | Plurality With Runoff | 0.82 | 0.92 | 1.00 | 1.00 | n/a | n/a | n/a |
| 4 | 100 | 0.80 | Schulze | 0.82 | 0.91 | 1.00 | 1.00 | n/a | n/a | n/a |
| 4 | 100 | 0.90 | Bucklin | 0.54 | 1.00 | 1.00 | 1.00 | n/a | n/a | n/a |
| 4 | 100 | 0.90 | Coombs | 0.59 | 0.73 | 1.00 | 1.00 | n/a | n/a | n/a |
| 4 | 100 | 0.90 | Plurality With Runoff | 0.66 | 0.82 | 1.00 | 1.00 | n/a | n/a | n/a |
| 4 | 100 | 0.90 | Schulze | 0.63 | 0.78 | 1.00 | 1.00 | n/a | n/a | n/a |
| 4 | 100 | 1.00 | Bucklin | 0.44 | 1.00 | 1.00 | 1.00 | n/a | n/a | n/a |
| 4 | 100 | 1.00 | Coombs | 0.48 | 0.64 | 1.00 | 1.00 | n/a | n/a | n/a |
| 4 | 100 | 1.00 | Plurality With Runoff | 0.61 | 0.76 | 1.00 | 1.00 | n/a | n/a | n/a |
| 4 | 100 | 1.00 | Schulze | 0.55 | 0.72 | 1.00 | 1.00 | n/a | n/a | n/a |
| 4 | 200 | 0.70 | Bucklin | 0.95 | 1.00 | 1.00 | 1.00 | n/a | n/a | n/a |
| 4 | 200 | 0.70 | Coombs | 0.97 | 0.99 | 1.00 | 1.00 | n/a | n/a | n/a |
| 4 | 200 | 0.70 | Plurality With Runoff | 0.97 | 0.99 | 1.00 | 1.00 | n/a | n/a | n/a |
| 4 | 200 | 0.70 | Schulze | 0.97 | 0.99 | 1.00 | 1.00 | n/a | n/a | n/a |
| 4 | 200 | 0.80 | Bucklin | 0.85 | 1.00 | 1.00 | 1.00 | n/a | n/a | n/a |
| 4 | 200 | 0.80 | Coombs | 0.88 | 0.95 | 1.00 | 1.00 | n/a | n/a | n/a |
| 4 | 200 | 0.80 | Plurality With Runoff | 0.89 | 0.95 | 1.00 | 1.00 | n/a | n/a | n/a |
| 4 | 200 | 0.80 | Schulze | 0.89 | 0.95 | 1.00 | 1.00 | n/a | n/a | n/a |
| 4 | 200 | 0.90 | Bucklin | 0.65 | 1.00 | 1.00 | 1.00 | n/a | n/a | n/a |
| 4 | 200 | 0.90 | Coombs | 0.67 | 0.82 | 1.00 | 1.00 | n/a | n/a | n/a |
| 4 | 200 | 0.90 | Plurality With Runoff | 0.71 | 0.87 | 1.00 | 1.00 | n/a | n/a | n/a |
| 4 | 200 | 0.90 | Schulze | 0.69 | 0.85 | 1.00 | 1.00 | n/a | n/a | n/a |
| 4 | 200 | 1.00 | Bucklin | 0.48 | 1.00 | 1.00 | 1.00 | n/a | n/a | n/a |
| 4 | 200 | 1.00 | Coombs | 0.52 | 0.65 | 1.00 | 1.00 | n/a | n/a | n/a |
| 4 | 200 | 1.00 | Plurality With Runoff | 0.65 | 0.77 | 1.00 | 1.00 | n/a | n/a | n/a |

| 4 | 200 | 1.00 | Schulze | 0.58 | 0.73 | 1.00 | 1.00 | n/a | n/a | n/a |
|---|-----|------|---------|------|------|------|------|-----|-----|-----|
| 4 | 300 | 0.70 | Bucklin | 0.98 | 1.00 | 1.00 | 1.00 | n/a | n/a | n/a |
| 4 | 300 | 0.70 | Coombs | 0.99 | 1.00 | 1.00 | 1.00 | n/a | n/a | n/a |
| 4 | 300 | 0.70 | Plurality With Runoff | 0.99 | 1.00 | 1.00 | 1.00 | n/a | n/a | n/a |
| 4 | 300 | 0.70 | Schulze | 0.99 | 1.00 | 1.00 | 1.00 | n/a | n/a | n/a |
| 4 | 300 | 0.80 | Bucklin | 0.91 | 1.00 | 1.00 | 1.00 | n/a | n/a | n/a |
| 4 | 300 | 0.80 | Coombs | 0.94 | 0.98 | 1.00 | 1.00 | n/a | n/a | n/a |
| 4 | 300 | 0.80 | Plurality With Runoff | 0.94 | 0.98 | 1.00 | 1.00 | n/a | n/a | n/a |
| 4 | 300 | 0.80 | Schulze | 0.94 | 0.98 | 1.00 | 1.00 | n/a | n/a | n/a |
| 4 | 300 | 0.90 | Bucklin | 0.67 | 1.00 | 1.00 | 1.00 | n/a | n/a | n/a |
| 4 | 300 | 0.90 | Coombs | 0.73 | 0.87 | 1.00 | 1.00 | n/a | n/a | n/a |
| 4 | 300 | 0.90 | Plurality With Runoff | 0.77 | 0.90 | 1.00 | 1.00 | n/a | n/a | n/a |
| 4 | 300 | 0.90 | Schulze | 0.75 | 0.90 | 1.00 | 1.00 | n/a | n/a | n/a |
| 4 | 300 | 1.00 | Bucklin | 0.45 | 1.00 | 1.00 | 1.00 | n/a | n/a | n/a |
| 4 | 300 | 1.00 | Coombs | 0.50 | 0.64 | 1.00 | 1.00 | n/a | n/a | n/a |
| 4 | 300 | 1.00 | Plurality With Runoff | 0.63 | 0.78 | 1.00 | 1.00 | n/a | n/a | n/a |
| 4 | 300 | 1.00 | Schulze | 0.56 | 0.72 | 1.00 | 1.00 | n/a | n/a | n/a |
| 4 | 400 | 0.70 | Bucklin | 1.00 | 1.00 | 1.00 | 1.00 | n/a | n/a | n/a |
| 4 | 400 | 0.70 | Coombs | 1.00 | 1.00 | 1.00 | 1.00 | n/a | n/a | n/a |
| 4 | 400 | 0.70 | Plurality With Runoff | 1.00 | 1.00 | 1.00 | 1.00 | n/a | n/a | n/a |
| 4 | 400 | 0.70 | Schulze | 1.00 | 1.00 | 1.00 | 1.00 | n/a | n/a | n/a |
| 4 | 400 | 0.80 | Bucklin | 0.94 | 1.00 | 1.00 | 1.00 | n/a | n/a | n/a |
| 4 | 400 | 0.80 | Coombs | 0.96 | 0.99 | 1.00 | 1.00 | n/a | n/a | n/a |
| 4 | 400 | 0.80 | Plurality With Runoff | 0.96 | 0.99 | 1.00 | 1.00 | n/a | n/a | n/a |
| 4 | 400 | 0.80 | Schulze | 0.96 | 0.99 | 1.00 | 1.00 | n/a | n/a | n/a |
| 4 | 400 | 0.90 | Bucklin | 0.73 | 1.00 | 1.00 | 1.00 | n/a | n/a | n/a |
| 4 | 400 | 0.90 | Coombs | 0.77 | 0.88 | 1.00 | 1.00 | n/a | n/a | n/a |
| 4 | 400 | 0.90 | Plurality With Runoff | 0.81 | 0.91 | 1.00 | 1.00 | n/a | n/a | n/a |
| 4 | 400 | 0.90 | Schulze | 0.79 | 0.90 | 1.00 | 1.00 | n/a | n/a | n/a |
| 4 | 400 | 1.00 | Bucklin | 0.44 | 1.00 | 1.00 | 1.00 | n/a | n/a | n/a |
| 4 | 400 | 1.00 | Coombs | 0.49 | 0.66 | 1.00 | 1.00 | n/a | n/a | n/a |
| 4 | 400 | 1.00 | Plurality With Runoff | 0.61 | 0.80 | 1.00 | 1.00 | n/a | n/a | n/a |
| 4 | 400 | 1.00 | Schulze | 0.55 | 0.76 | 1.00 | 1.00 | n/a | n/a | n/a |
| 4 | 500 | 0.70 | Bucklin | 1.00 | 1.00 | 1.00 | 1.00 | n/a | n/a | n/a |
| 4 | 500 | 0.70 | Coombs | 1.00 | 1.00 | 1.00 | 1.00 | n/a | n/a | n/a |
| 4 | 500 | 0.70 | Plurality With Runoff | 1.00 | 1.00 | 1.00 | 1.00 | n/a | n/a | n/a |

| 4 | 500 | 0.70 | Schulze | 1.00 | 1.00 | 1.00 | 1.00 | n/a | n/a | n/a |
|---|---|---|---|---|---|---|---|---|---|---|
| 4 | 500 | 0.80 | Bucklin | 0.96 | 1.00 | 1.00 | 1.00 | n/a | n/a | n/a |
| 4 | 500 | 0.80 | Coombs | 0.97 | 0.99 | 1.00 | 1.00 | n/a | n/a | n/a |
| 4 | 500 | 0.80 | Plurality With Runoff | 0.97 | 0.99 | 1.00 | 1.00 | n/a | n/a | n/a |
| 4 | 500 | 0.80 | Schulze | 0.97 | 0.99 | 1.00 | 1.00 | n/a | n/a | n/a |
| 4 | 500 | 0.90 | Bucklin | 0.77 | 1.00 | 1.00 | 1.00 | n/a | n/a | n/a |
| 4 | 500 | 0.90 | Coombs | 0.80 | 0.91 | 1.00 | 1.00 | n/a | n/a | n/a |
| 4 | 500 | 0.90 | Plurality With Runoff | 0.81 | 0.92 | 1.00 | 1.00 | n/a | n/a | n/a |
| 4 | 500 | 0.90 | Schulze | 0.80 | 0.91 | 1.00 | 1.00 | n/a | n/a | n/a |
| 4 | 500 | 1.00 | Bucklin | 0.48 | 1.00 | 1.00 | 1.00 | n/a | n/a | n/a |
| 4 | 500 | 1.00 | Coombs | 0.50 | 0.65 | 1.00 | 1.00 | n/a | n/a | n/a |
| 4 | 500 | 1.00 | Plurality With Runoff | 0.64 | 0.82 | 1.00 | 1.00 | n/a | n/a | n/a |
| 4 | 500 | 1.00 | Schulze | 0.57 | 0.76 | 1.00 | 1.00 | n/a | n/a | n/a |
| 4 | 600 | 0.70 | Bucklin | 1.00 | 1.00 | 1.00 | 1.00 | n/a | n/a | n/a |
| 4 | 600 | 0.70 | Coombs | 1.00 | 1.00 | 1.00 | 1.00 | n/a | n/a | n/a |
| 4 | 600 | 0.70 | Plurality With Runoff | 1.00 | 1.00 | 1.00 | 1.00 | n/a | n/a | n/a |
| 4 | 600 | 0.70 | Schulze | 1.00 | 1.00 | 1.00 | 1.00 | n/a | n/a | n/a |
| 4 | 600 | 0.80 | Bucklin | 0.97 | 1.00 | 1.00 | 1.00 | n/a | n/a | n/a |
| 4 | 600 | 0.80 | Coombs | 0.98 | 1.00 | 1.00 | 1.00 | n/a | n/a | n/a |
| 4 | 600 | 0.80 | Plurality With Runoff | 0.98 | 1.00 | 1.00 | 1.00 | n/a | n/a | n/a |
| 4 | 600 | 0.80 | Schulze | 0.98 | 1.00 | 1.00 | 1.00 | n/a | n/a | n/a |
| 4 | 600 | 0.90 | Bucklin | 0.80 | 1.00 | 1.00 | 1.00 | n/a | n/a | n/a |
| 4 | 600 | 0.90 | Coombs | 0.85 | 0.92 | 1.00 | 1.00 | n/a | n/a | n/a |
| 4 | 600 | 0.90 | Plurality With Runoff | 0.86 | 0.93 | 1.00 | 1.00 | n/a | n/a | n/a |
| 4 | 600 | 0.90 | Schulze | 0.85 | 0.93 | 1.00 | 1.00 | n/a | n/a | n/a |
| 4 | 600 | 1.00 | Bucklin | 0.48 | 1.00 | 1.00 | 1.00 | n/a | n/a | n/a |
| 4 | 600 | 1.00 | Coombs | 0.49 | 0.66 | 1.00 | 1.00 | n/a | n/a | n/a |
| 4 | 600 | 1.00 | Plurality With Runoff | 0.63 | 0.81 | 1.00 | 1.00 | n/a | n/a | n/a |
| 4 | 600 | 1.00 | Schulze | 0.57 | 0.76 | 1.00 | 1.00 | n/a | n/a | n/a |
| 4 | 2,000 | 0.70 | Bucklin | 1.00 | 1.00 | 1.00 | 1.00 | n/a | n/a | n/a |
| 4 | 2,000 | 0.70 | Coombs | 1.00 | 1.00 | 1.00 | 1.00 | n/a | n/a | n/a |
| 4 | 2,000 | 0.70 | Plurality With Runoff | 1.00 | 1.00 | 1.00 | 1.00 | n/a | n/a | n/a |
| 4 | 2,000 | 0.70 | Schulze | 1.00 | 1.00 | 1.00 | 1.00 | n/a | n/a | n/a |
| 4 | 2,000 | 0.80 | Bucklin | 1.00 | 1.00 | 1.00 | 1.00 | n/a | n/a | n/a |
| 4 | 2,000 | 0.80 | Coombs | 1.00 | 1.00 | 1.00 | 1.00 | n/a | n/a | n/a |
| 4 | 2,000 | 0.80 | Plurality With Runoff | 1.00 | 1.00 | 1.00 | 1.00 | n/a | n/a | n/a |

| 4 | 2,000 | 0.80 | Schulze | 1.00 | 1.00 | 1.00 | 1.00 | n/a | n/a | n/a |
| 4 | 2,000 | 0.90 | Bucklin | 0.94 | 1.00 | 1.00 | 1.00 | n/a | n/a | n/a |
| 4 | 2,000 | 0.90 | Coombs | 0.96 | 0.99 | 1.00 | 1.00 | n/a | n/a | n/a |
| 4 | 2,000 | 0.90 | Plurality With Runoff | 0.96 | 0.99 | 1.00 | 1.00 | n/a | n/a | n/a |
| 4 | 2,000 | 0.90 | Schulze | 0.96 | 0.99 | 1.00 | 1.00 | n/a | n/a | n/a |
| 4 | 2,000 | 1.00 | Bucklin | 0.47 | 1.00 | 1.00 | 1.00 | n/a | n/a | n/a |
| 4 | 2,000 | 1.00 | Coombs | 0.49 | 0.67 | 1.00 | 1.00 | n/a | n/a | n/a |
| 4 | 2,000 | 1.00 | Plurality With Runoff | 0.65 | 0.83 | 1.00 | 1.00 | n/a | n/a | n/a |
| 4 | 2,000 | 1.00 | Schulze | 0.57 | 0.76 | 1.00 | 1.00 | n/a | n/a | n/a |
| 5 | 100 | 0.70 | Bucklin | 0.88 | 1.00 | 1.00 | 1.00 | 1.00 | n/a | n/a |
| 5 | 100 | 0.70 | Coombs | 0.92 | 0.97 | 0.98 | 1.00 | 1.00 | n/a | n/a |
| 5 | 100 | 0.70 | Plurality With Runoff | 0.92 | 0.97 | 0.98 | 1.00 | 1.00 | n/a | n/a |
| 5 | 100 | 0.70 | Schulze | 0.92 | 0.97 | 0.99 | 1.00 | 1.00 | n/a | n/a |
| 5 | 100 | 0.80 | Bucklin | 0.75 | 0.91 | 1.00 | 1.00 | 1.00 | n/a | n/a |
| 5 | 100 | 0.80 | Coombs | 0.79 | 0.87 | 0.93 | 1.00 | 1.00 | n/a | n/a |
| 5 | 100 | 0.80 | Plurality With Runoff | 0.82 | 0.90 | 0.94 | 1.00 | 1.00 | n/a | n/a |
| 5 | 100 | 0.80 | Schulze | 0.80 | 0.89 | 0.94 | 1.00 | 1.00 | n/a | n/a |
| 5 | 100 | 0.90 | Bucklin | 0.50 | 0.63 | 1.00 | 1.00 | 1.00 | n/a | n/a |
| 5 | 100 | 0.90 | Coombs | 0.56 | 0.68 | 0.77 | 1.00 | 1.00 | n/a | n/a |
| 5 | 100 | 0.90 | Plurality With Runoff | 0.65 | 0.78 | 0.86 | 1.00 | 1.00 | n/a | n/a |
| 5 | 100 | 0.90 | Schulze | 0.60 | 0.71 | 0.82 | 1.00 | 1.00 | n/a | n/a |
| 5 | 100 | 1.00 | Bucklin | 0.32 | 0.48 | 1.00 | 1.00 | 1.00 | n/a | n/a |
| 5 | 100 | 1.00 | Coombs | 0.38 | 0.52 | 0.64 | 1.00 | 1.00 | n/a | n/a |
| 5 | 100 | 1.00 | Plurality With Runoff | 0.55 | 0.70 | 0.83 | 1.00 | 1.00 | n/a | n/a |
| 5 | 100 | 1.00 | Schulze | 0.46 | 0.62 | 0.78 | 1.00 | 1.00 | n/a | n/a |
| 5 | 200 | 0.70 | Bucklin | 0.96 | 1.00 | 1.00 | 1.00 | 1.00 | n/a | n/a |
| 5 | 200 | 0.70 | Coombs | 0.98 | 0.99 | 0.99 | 1.00 | 1.00 | n/a | n/a |
| 5 | 200 | 0.70 | Plurality With Runoff | 0.98 | 0.99 | 0.99 | 1.00 | 1.00 | n/a | n/a |
| 5 | 200 | 0.70 | Schulze | 0.98 | 0.99 | 0.99 | 1.00 | 1.00 | n/a | n/a |
| 5 | 200 | 0.80 | Bucklin | 0.82 | 0.96 | 1.00 | 1.00 | 1.00 | n/a | n/a |
| 5 | 200 | 0.80 | Coombs | 0.85 | 0.93 | 0.96 | 1.00 | 1.00 | n/a | n/a |
| 5 | 200 | 0.80 | Plurality With Runoff | 0.85 | 0.94 | 0.96 | 1.00 | 1.00 | n/a | n/a |
| 5 | 200 | 0.80 | Schulze | 0.85 | 0.94 | 0.96 | 1.00 | 1.00 | n/a | n/a |
| 5 | 200 | 0.90 | Bucklin | 0.55 | 0.67 | 1.00 | 1.00 | 1.00 | n/a | n/a |
| 5 | 200 | 0.90 | Coombs | 0.61 | 0.76 | 0.85 | 1.00 | 1.00 | n/a | n/a |
| 5 | 200 | 0.90 | Plurality With Runoff | 0.69 | 0.84 | 0.91 | 1.00 | 1.00 | n/a | n/a |

| 5 | 200 | 0.90 | Schulze | 0.64 | 0.80 | 0.88 | 1.00 | 1.00 | n/a | n/a |
|---|-----|------|---------|------|------|------|------|------|-----|-----|
| 5 | 200 | 1.00 | Bucklin | 0.32 | 0.46 | 1.00 | 1.00 | 1.00 | n/a | n/a |
| 5 | 200 | 1.00 | Coombs | 0.38 | 0.50 | 0.65 | 1.00 | 1.00 | n/a | n/a |
| 5 | 200 | 1.00 | Plurality With Runoff | 0.57 | 0.73 | 0.85 | 1.00 | 1.00 | n/a | n/a |
| 5 | 200 | 1.00 | Schulze | 0.47 | 0.61 | 0.77 | 1.00 | 1.00 | n/a | n/a |
| 5 | 300 | 0.70 | Bucklin | 0.99 | 1.00 | 1.00 | 1.00 | 1.00 | n/a | n/a |
| 5 | 300 | 0.70 | Coombs | 0.99 | 1.00 | 1.00 | 1.00 | 1.00 | n/a | n/a |
| 5 | 300 | 0.70 | Plurality With Runoff | 0.99 | 1.00 | 1.00 | 1.00 | 1.00 | n/a | n/a |
| 5 | 300 | 0.70 | Schulze | 0.99 | 1.00 | 1.00 | 1.00 | 1.00 | n/a | n/a |
| 5 | 300 | 0.80 | Bucklin | 0.88 | 0.97 | 1.00 | 1.00 | 1.00 | n/a | n/a |
| 5 | 300 | 0.80 | Coombs | 0.91 | 0.97 | 0.99 | 1.00 | 1.00 | n/a | n/a |
| 5 | 300 | 0.80 | Plurality With Runoff | 0.91 | 0.97 | 0.99 | 1.00 | 1.00 | n/a | n/a |
| 5 | 300 | 0.80 | Schulze | 0.91 | 0.97 | 0.99 | 1.00 | 1.00 | n/a | n/a |
| 5 | 300 | 0.90 | Bucklin | 0.61 | 0.74 | 1.00 | 1.00 | 1.00 | n/a | n/a |
| 5 | 300 | 0.90 | Coombs | 0.68 | 0.81 | 0.88 | 1.00 | 1.00 | n/a | n/a |
| 5 | 300 | 0.90 | Plurality With Runoff | 0.73 | 0.86 | 0.92 | 1.00 | 1.00 | n/a | n/a |
| 5 | 300 | 0.90 | Schulze | 0.70 | 0.84 | 0.91 | 1.00 | 1.00 | n/a | n/a |
| 5 | 300 | 1.00 | Bucklin | 0.33 | 0.46 | 1.00 | 1.00 | 1.00 | n/a | n/a |
| 5 | 300 | 1.00 | Coombs | 0.41 | 0.55 | 0.66 | 1.00 | 1.00 | n/a | n/a |
| 5 | 300 | 1.00 | Plurality With Runoff | 0.59 | 0.71 | 0.85 | 1.00 | 1.00 | n/a | n/a |
| 5 | 300 | 1.00 | Schulze | 0.49 | 0.64 | 0.77 | 1.00 | 1.00 | n/a | n/a |
| 5 | 400 | 0.70 | Bucklin | 1.00 | 1.00 | 1.00 | 1.00 | 1.00 | n/a | n/a |
| 5 | 400 | 0.70 | Coombs | 1.00 | 1.00 | 1.00 | 1.00 | 1.00 | n/a | n/a |
| 5 | 400 | 0.70 | Plurality With Runoff | 1.00 | 1.00 | 1.00 | 1.00 | 1.00 | n/a | n/a |
| 5 | 400 | 0.70 | Schulze | 1.00 | 1.00 | 1.00 | 1.00 | 1.00 | n/a | n/a |
| 5 | 400 | 0.80 | Bucklin | 0.93 | 0.98 | 1.00 | 1.00 | 1.00 | n/a | n/a |
| 5 | 400 | 0.80 | Coombs | 0.95 | 0.98 | 0.99 | 1.00 | 1.00 | n/a | n/a |
| 5 | 400 | 0.80 | Plurality With Runoff | 0.95 | 0.98 | 0.99 | 1.00 | 1.00 | n/a | n/a |
| 5 | 400 | 0.80 | Schulze | 0.95 | 0.98 | 0.99 | 1.00 | 1.00 | n/a | n/a |
| 5 | 400 | 0.90 | Bucklin | 0.66 | 0.75 | 1.00 | 1.00 | 1.00 | n/a | n/a |
| 5 | 400 | 0.90 | Coombs | 0.73 | 0.82 | 0.91 | 1.00 | 1.00 | n/a | n/a |
| 5 | 400 | 0.90 | Plurality With Runoff | 0.78 | 0.86 | 0.93 | 1.00 | 1.00 | n/a | n/a |
| 5 | 400 | 0.90 | Schulze | 0.75 | 0.84 | 0.93 | 1.00 | 1.00 | n/a | n/a |
| 5 | 400 | 1.00 | Bucklin | 0.32 | 0.47 | 1.00 | 1.00 | 1.00 | n/a | n/a |
| 5 | 400 | 1.00 | Coombs | 0.41 | 0.51 | 0.65 | 1.00 | 1.00 | n/a | n/a |
| 5 | 400 | 1.00 | Plurality With Runoff | 0.58 | 0.73 | 0.83 | 1.00 | 1.00 | n/a | n/a |

| 5 | 400 | 1.00 | Schulze | 0.48 | 0.61 | 0.75 | 1.00 | 1.00 | n/a | n/a |
|---|-----|------|---------|------|------|------|------|------|-----|-----|
| 5 | 500 | 0.70 | Bucklin | 1.00 | 1.00 | 1.00 | 1.00 | 1.00 | n/a | n/a |
| 5 | 500 | 0.70 | Coombs | 1.00 | 1.00 | 1.00 | 1.00 | 1.00 | n/a | n/a |
| 5 | 500 | 0.70 | Plurality With Runoff | 1.00 | 1.00 | 1.00 | 1.00 | 1.00 | n/a | n/a |
| 5 | 500 | 0.70 | Schulze | 1.00 | 1.00 | 1.00 | 1.00 | 1.00 | n/a | n/a |
| 5 | 500 | 0.80 | Bucklin | 0.95 | 1.00 | 1.00 | 1.00 | 1.00 | n/a | n/a |
| 5 | 500 | 0.80 | Coombs | 0.96 | 0.99 | 0.99 | 1.00 | 1.00 | n/a | n/a |
| 5 | 500 | 0.80 | Plurality With Runoff | 0.96 | 0.99 | 0.99 | 1.00 | 1.00 | n/a | n/a |
| 5 | 500 | 0.80 | Schulze | 0.96 | 0.99 | 0.99 | 1.00 | 1.00 | n/a | n/a |
| 5 | 500 | 0.90 | Bucklin | 0.73 | 0.79 | 1.00 | 1.00 | 1.00 | n/a | n/a |
| 5 | 500 | 0.90 | Coombs | 0.78 | 0.87 | 0.93 | 1.00 | 1.00 | n/a | n/a |
| 5 | 500 | 0.90 | Plurality With Runoff | 0.80 | 0.89 | 0.94 | 1.00 | 1.00 | n/a | n/a |
| 5 | 500 | 0.90 | Schulze | 0.79 | 0.88 | 0.94 | 1.00 | 1.00 | n/a | n/a |
| 5 | 500 | 1.00 | Bucklin | 0.34 | 0.45 | 1.00 | 1.00 | 1.00 | n/a | n/a |
| 5 | 500 | 1.00 | Coombs | 0.42 | 0.53 | 0.67 | 1.00 | 1.00 | n/a | n/a |
| 5 | 500 | 1.00 | Plurality With Runoff | 0.62 | 0.74 | 0.85 | 1.00 | 1.00 | n/a | n/a |
| 5 | 500 | 1.00 | Schulze | 0.50 | 0.63 | 0.80 | 1.00 | 1.00 | n/a | n/a |
| 5 | 600 | 0.70 | Bucklin | 1.00 | 1.00 | 1.00 | 1.00 | 1.00 | n/a | n/a |
| 5 | 600 | 0.70 | Coombs | 1.00 | 1.00 | 1.00 | 1.00 | 1.00 | n/a | n/a |
| 5 | 600 | 0.70 | Plurality With Runoff | 1.00 | 1.00 | 1.00 | 1.00 | 1.00 | n/a | n/a |
| 5 | 600 | 0.70 | Schulze | 1.00 | 1.00 | 1.00 | 1.00 | 1.00 | n/a | n/a |
| 5 | 600 | 0.80 | Bucklin | 0.97 | 1.00 | 1.00 | 1.00 | 1.00 | n/a | n/a |
| 5 | 600 | 0.80 | Coombs | 0.98 | 1.00 | 1.00 | 1.00 | 1.00 | n/a | n/a |
| 5 | 600 | 0.80 | Plurality With Runoff | 0.98 | 1.00 | 1.00 | 1.00 | 1.00 | n/a | n/a |
| 5 | 600 | 0.80 | Schulze | 0.98 | 1.00 | 1.00 | 1.00 | 1.00 | n/a | n/a |
| 5 | 600 | 0.90 | Bucklin | 0.72 | 0.81 | 1.00 | 1.00 | 1.00 | n/a | n/a |
| 5 | 600 | 0.90 | Coombs | 0.80 | 0.90 | 0.94 | 1.00 | 1.00 | n/a | n/a |
| 5 | 600 | 0.90 | Plurality With Runoff | 0.81 | 0.91 | 0.94 | 1.00 | 1.00 | n/a | n/a |
| 5 | 600 | 0.90 | Schulze | 0.81 | 0.90 | 0.94 | 1.00 | 1.00 | n/a | n/a |
| 5 | 600 | 1.00 | Bucklin | 0.32 | 0.46 | 1.00 | 1.00 | 1.00 | n/a | n/a |
| 5 | 600 | 1.00 | Coombs | 0.39 | 0.52 | 0.65 | 1.00 | 1.00 | n/a | n/a |
| 5 | 600 | 1.00 | Plurality With Runoff | 0.59 | 0.74 | 0.86 | 1.00 | 1.00 | n/a | n/a |
| 5 | 600 | 1.00 | Schulze | 0.49 | 0.63 | 0.78 | 1.00 | 1.00 | n/a | n/a |
| 5 | 2,000 | 0.70 | Bucklin | 1.00 | 1.00 | 1.00 | 1.00 | 1.00 | n/a | n/a |
| 5 | 2,000 | 0.70 | Coombs | 1.00 | 1.00 | 1.00 | 1.00 | 1.00 | n/a | n/a |
| 5 | 2,000 | 0.70 | Plurality With Runoff | 1.00 | 1.00 | 1.00 | 1.00 | 1.00 | n/a | n/a |
| 5 | 2,000 | 0.70 | Schulze | 1.00 | 1.00 | 1.00 | 1.00 | 1.00 | n/a | n/a |
| 5 | 2,000 | 0.80 | Bucklin | 1.00 | 1.00 | 1.00 | 1.00 | 1.00 | n/a | n/a |
| 5 | 2,000 | 0.80 | Coombs | 1.00 | 1.00 | 1.00 | 1.00 | 1.00 | n/a | n/a |
| 5 | 2,000 | 0.80 | Plurality With Runoff | 1.00 | 1.00 | 1.00 | 1.00 | 1.00 | n/a | n/a |
| 5 | 2,000 | 0.80 | Schulze | 1.00 | 1.00 | 1.00 | 1.00 | 1.00 | n/a | n/a |
| 5 | 2,000 | 0.90 | Bucklin | 0.91 | 0.94 | 1.00 | 1.00 | 1.00 | n/a | n/a |
| 5 | 2,000 | 0.90 | Coombs | 0.94 | 0.98 | 0.99 | 1.00 | 1.00 | n/a | n/a |
| 5 | 2,000 | 0.90 | Plurality With Runoff | 0.94 | 0.98 | 0.99 | 1.00 | 1.00 | n/a | n/a |
| 5 | 2,000 | 0.90 | Schulze | 0.94 | 0.98 | 0.99 | 1.00 | 1.00 | n/a | n/a |
| 5 | 2,000 | 1.00 | Bucklin | 0.37 | 0.50 | 1.00 | 1.00 | 1.00 | n/a | n/a |
| 5 | 2,000 | 1.00 | Coombs | 0.42 | 0.56 | 0.67 | 1.00 | 1.00 | n/a | n/a |
| 5 | 2,000 | 1.00 | Plurality With Runoff | 0.61 | 0.76 | 0.86 | 1.00 | 1.00 | n/a | n/a |
| 5 | 2,000 | 1.00 | Schulze | 0.49 | 0.64 | 0.77 | 1.00 | 1.00 | n/a | n/a |
| 6 | 100 | 0.70 | Bucklin | 0.87 | 0.98 | 1.00 | 1.00 | 1.00 | 1.00 | n/a |
| 6 | 100 | 0.70 | Coombs | 0.90 | 0.96 | 0.97 | 0.98 | 1.00 | 1.00 | n/a |
| 6 | 100 | 0.70 | Plurality With Runoff | 0.91 | 0.96 | 0.98 | 0.99 | 1.00 | 1.00 | n/a |
| 6 | 100 | 0.70 | Schulze | 0.91 | 0.96 | 0.97 | 0.98 | 1.00 | 1.00 | n/a |
| 6 | 100 | 0.80 | Bucklin | 0.71 | 0.82 | 1.00 | 1.00 | 1.00 | 1.00 | n/a |
| 6 | 100 | 0.80 | Coombs | 0.76 | 0.86 | 0.91 | 0.96 | 1.00 | 1.00 | n/a |
| 6 | 100 | 0.80 | Plurality With Runoff | 0.78 | 0.87 | 0.93 | 0.97 | 1.00 | 1.00 | n/a |
| 6 | 100 | 0.80 | Schulze | 0.76 | 0.86 | 0.92 | 0.97 | 1.00 | 1.00 | n/a |
| 6 | 100 | 0.90 | Bucklin | 0.46 | 0.62 | 1.00 | 1.00 | 1.00 | 1.00 | n/a |
| 6 | 100 | 0.90 | Coombs | 0.50 | 0.63 | 0.71 | 0.82 | 1.00 | 1.00 | n/a |
| 6 | 100 | 0.90 | Plurality With Runoff | 0.61 | 0.76 | 0.84 | 0.91 | 1.00 | 1.00 | n/a |
| 6 | 100 | 0.90 | Schulze | 0.53 | 0.69 | 0.77 | 0.87 | 1.00 | 1.00 | n/a |
| 6 | 100 | 1.00 | Bucklin | 0.30 | 0.45 | 1.00 | 1.00 | 1.00 | 1.00 | n/a |
| 6 | 100 | 1.00 | Coombs | 0.32 | 0.43 | 0.53 | 0.63 | 1.00 | 1.00 | n/a |
| 6 | 100 | 1.00 | Plurality With Runoff | 0.47 | 0.64 | 0.73 | 0.83 | 1.00 | 1.00 | n/a |
| 6 | 100 | 1.00 | Schulze | 0.38 | 0.51 | 0.65 | 0.76 | 1.00 | 1.00 | n/a |
| 6 | 200 | 0.70 | Bucklin | 0.95 | 1.00 | 1.00 | 1.00 | 1.00 | 1.00 | n/a |
| 6 | 200 | 0.70 | Coombs | 0.96 | 0.99 | 1.00 | 1.00 | 1.00 | 1.00 | n/a |
| 6 | 200 | 0.70 | Plurality With Runoff | 0.96 | 0.99 | 1.00 | 1.00 | 1.00 | 1.00 | n/a |
| 6 | 200 | 0.70 | Schulze | 0.96 | 0.99 | 1.00 | 1.00 | 1.00 | 1.00 | n/a |
| 6 | 200 | 0.80 | Bucklin | 0.81 | 0.89 | 1.00 | 1.00 | 1.00 | 1.00 | n/a |
| 6 | 200 | 0.80 | Coombs | 0.86 | 0.94 | 0.95 | 0.98 | 1.00 | 1.00 | n/a |
| 6 | 200 | 0.80 | Plurality With Runoff | 0.86 | 0.94 | 0.95 | 0.99 | 1.00 | 1.00 | n/a |
| 6 | 200 | 0.80 | Schulze | 0.86 | 0.94 | 0.95 | 0.98 | 1.00 | 1.00 | n/a |
| 6 | 200 | 0.90 | Bucklin | 0.57 | 0.69 | 1.00 | 1.00 | 1.00 | 1.00 | n/a |
| 6 | 200 | 0.90 | Coombs | 0.60 | 0.71 | 0.80 | 0.86 | 1.00 | 1.00 | n/a |
| 6 | 200 | 0.90 | Plurality With Runoff | 0.68 | 0.79 | 0.87 | 0.92 | 1.00 | 1.00 | n/a |
| 6 | 200 | 0.90 | Schulze | 0.63 | 0.76 | 0.84 | 0.90 | 1.00 | 1.00 | n/a |
| 6 | 200 | 1.00 | Bucklin | 0.32 | 0.45 | 1.00 | 1.00 | 1.00 | 1.00 | n/a |
| 6 | 200 | 1.00 | Coombs | 0.37 | 0.46 | 0.55 | 0.65 | 1.00 | 1.00 | n/a |
| 6 | 200 | 1.00 | Plurality With Runoff | 0.53 | 0.66 | 0.75 | 0.85 | 1.00 | 1.00 | n/a |
| 6 | 200 | 1.00 | Schulze | 0.41 | 0.55 | 0.65 | 0.79 | 1.00 | 1.00 | n/a |
| 6 | 300 | 0.70 | Bucklin | 0.98 | 1.00 | 1.00 | 1.00 | 1.00 | 1.00 | n/a |
| 6 | 300 | 0.70 | Coombs | 0.99 | 1.00 | 1.00 | 1.00 | 1.00 | 1.00 | n/a |
| 6 | 300 | 0.70 | Plurality With Runoff | 0.99 | 1.00 | 1.00 | 1.00 | 1.00 | 1.00 | n/a |
| 6 | 300 | 0.70 | Schulze | 0.99 | 1.00 | 1.00 | 1.00 | 1.00 | 1.00 | n/a |
| 6 | 300 | 0.80 | Bucklin | 0.88 | 0.92 | 1.00 | 1.00 | 1.00 | 1.00 | n/a |
| 6 | 300 | 0.80 | Coombs | 0.91 | 0.97 | 0.98 | 0.99 | 1.00 | 1.00 | n/a |
| 6 | 300 | 0.80 | Plurality With Runoff | 0.91 | 0.97 | 0.98 | 0.99 | 1.00 | 1.00 | n/a |
| 6 | 300 | 0.80 | Schulze | 0.91 | 0.97 | 0.98 | 0.99 | 1.00 | 1.00 | n/a |
| 6 | 300 | 0.90 | Bucklin | 0.61 | 0.74 | 1.00 | 1.00 | 1.00 | 1.00 | n/a |
| 6 | 300 | 0.90 | Coombs | 0.67 | 0.78 | 0.85 | 0.90 | 1.00 | 1.00 | n/a |
| 6 | 300 | 0.90 | Plurality With Runoff | 0.72 | 0.82 | 0.90 | 0.93 | 1.00 | 1.00 | n/a |
| 6 | 300 | 0.90 | Schulze | 0.68 | 0.79 | 0.88 | 0.92 | 1.00 | 1.00 | n/a |
| 6 | 300 | 1.00 | Bucklin | 0.31 | 0.45 | 1.00 | 1.00 | 1.00 | 1.00 | n/a |
| 6 | 300 | 1.00 | Coombs | 0.32 | 0.43 | 0.51 | 0.63 | 1.00 | 1.00 | n/a |
| 6 | 300 | 1.00 | Plurality With Runoff | 0.56 | 0.69 | 0.79 | 0.87 | 1.00 | 1.00 | n/a |
| 6 | 300 | 1.00 | Schulze | 0.40 | 0.55 | 0.64 | 0.78 | 1.00 | 1.00 | n/a |
| 6 | 400 | 0.70 | Bucklin | 0.99 | 1.00 | 1.00 | 1.00 | 1.00 | 1.00 | n/a |
| 6 | 400 | 0.70 | Coombs | 1.00 | 1.00 | 1.00 | 1.00 | 1.00 | 1.00 | n/a |
| 6 | 400 | 0.70 | Plurality With Runoff | 1.00 | 1.00 | 1.00 | 1.00 | 1.00 | 1.00 | n/a |
| 6 | 400 | 0.70 | Schulze | 1.00 | 1.00 | 1.00 | 1.00 | 1.00 | 1.00 | n/a |
| 6 | 400 | 0.80 | Bucklin | 0.92 | 0.95 | 1.00 | 1.00 | 1.00 | 1.00 | n/a |
| 6 | 400 | 0.80 | Coombs | 0.94 | 0.98 | 0.99 | 0.99 | 1.00 | 1.00 | n/a |
| 6 | 400 | 0.80 | Plurality With Runoff | 0.94 | 0.98 | 0.99 | 0.99 | 1.00 | 1.00 | n/a |
| 6 | 400 | 0.80 | Schulze | 0.94 | 0.98 | 0.99 | 0.99 | 1.00 | 1.00 | n/a |
| 6 | 400 | 0.90 | Bucklin | 0.63 | 0.74 | 1.00 | 1.00 | 1.00 | 1.00 | n/a |
| 6 | 400 | 0.90 | Coombs | 0.68 | 0.81 | 0.88 | 0.92 | 1.00 | 1.00 | n/a |
| 6 | 400 | 0.90 | Plurality With Runoff | 0.72 | 0.84 | 0.91 | 0.94 | 1.00 | 1.00 | n/a |

| 6 | 400 | 0.90 | Schulze | 0.70 | 0.82 | 0.89 | 0.93 | 1.00 | 1.00 | n/a |
|---|-----|------|---------|------|------|------|------|------|------|-----|
| 6 | 400 | 1.00 | Bucklin | 0.34 | 0.47 | 1.00 | 1.00 | 1.00 | 1.00 | n/a |
| 6 | 400 | 1.00 | Coombs | 0.35 | 0.44 | 0.53 | 0.66 | 1.00 | 1.00 | n/a |
| 6 | 400 | 1.00 | Plurality With Runoff | 0.56 | 0.71 | 0.80 | 0.90 | 1.00 | 1.00 | n/a |
| 6 | 400 | 1.00 | Schulze | 0.41 | 0.54 | 0.68 | 0.79 | 1.00 | 1.00 | n/a |
| 6 | 500 | 0.70 | Bucklin | 1.00 | 1.00 | 1.00 | 1.00 | 1.00 | 1.00 | n/a |
| 6 | 500 | 0.70 | Coombs | 1.00 | 1.00 | 1.00 | 1.00 | 1.00 | 1.00 | n/a |
| 6 | 500 | 0.70 | Plurality With Runoff | 1.00 | 1.00 | 1.00 | 1.00 | 1.00 | 1.00 | n/a |
| 6 | 500 | 0.70 | Schulze | 1.00 | 1.00 | 1.00 | 1.00 | 1.00 | 1.00 | n/a |
| 6 | 500 | 0.80 | Bucklin | 0.96 | 0.96 | 1.00 | 1.00 | 1.00 | 1.00 | n/a |
| 6 | 500 | 0.80 | Coombs | 0.97 | 0.99 | 0.99 | 1.00 | 1.00 | 1.00 | n/a |
| 6 | 500 | 0.80 | Plurality With Runoff | 0.97 | 0.99 | 0.99 | 1.00 | 1.00 | 1.00 | n/a |
| 6 | 500 | 0.80 | Schulze | 0.97 | 0.99 | 0.99 | 1.00 | 1.00 | 1.00 | n/a |
| 6 | 500 | 0.90 | Bucklin | 0.69 | 0.80 | 1.00 | 1.00 | 1.00 | 1.00 | n/a |
| 6 | 500 | 0.90 | Coombs | 0.75 | 0.86 | 0.91 | 0.94 | 1.00 | 1.00 | n/a |
| 6 | 500 | 0.90 | Plurality With Runoff | 0.78 | 0.89 | 0.94 | 0.96 | 1.00 | 1.00 | n/a |
| 6 | 500 | 0.90 | Schulze | 0.76 | 0.87 | 0.92 | 0.95 | 1.00 | 1.00 | n/a |
| 6 | 500 | 1.00 | Bucklin | 0.33 | 0.47 | 1.00 | 1.00 | 1.00 | 1.00 | n/a |
| 6 | 500 | 1.00 | Coombs | 0.31 | 0.41 | 0.50 | 0.64 | 1.00 | 1.00 | n/a |
| 6 | 500 | 1.00 | Plurality With Runoff | 0.54 | 0.69 | 0.79 | 0.86 | 1.00 | 1.00 | n/a |
| 6 | 500 | 1.00 | Schulze | 0.41 | 0.56 | 0.66 | 0.78 | 1.00 | 1.00 | n/a |
| 6 | 600 | 0.70 | Bucklin | 1.00 | 1.00 | 1.00 | 1.00 | 1.00 | 1.00 | n/a |
| 6 | 600 | 0.70 | Coombs | 1.00 | 1.00 | 1.00 | 1.00 | 1.00 | 1.00 | n/a |
| 6 | 600 | 0.70 | Plurality With Runoff | 1.00 | 1.00 | 1.00 | 1.00 | 1.00 | 1.00 | n/a |
| 6 | 600 | 0.70 | Schulze | 1.00 | 1.00 | 1.00 | 1.00 | 1.00 | 1.00 | n/a |
| 6 | 600 | 0.80 | Bucklin | 0.96 | 0.97 | 1.00 | 1.00 | 1.00 | 1.00 | n/a |
| 6 | 600 | 0.80 | Coombs | 0.97 | 0.99 | 1.00 | 1.00 | 1.00 | 1.00 | n/a |
| 6 | 600 | 0.80 | Plurality With Runoff | 0.97 | 0.99 | 1.00 | 1.00 | 1.00 | 1.00 | n/a |
| 6 | 600 | 0.80 | Schulze | 0.97 | 0.99 | 1.00 | 1.00 | 1.00 | 1.00 | n/a |
| 6 | 600 | 0.90 | Bucklin | 0.72 | 0.83 | 1.00 | 1.00 | 1.00 | 1.00 | n/a |
| 6 | 600 | 0.90 | Coombs | 0.77 | 0.88 | 0.93 | 0.95 | 1.00 | 1.00 | n/a |
| 6 | 600 | 0.90 | Plurality With Runoff | 0.79 | 0.90 | 0.94 | 0.96 | 1.00 | 1.00 | n/a |
| 6 | 600 | 0.90 | Schulze | 0.78 | 0.88 | 0.93 | 0.96 | 1.00 | 1.00 | n/a |
| 6 | 600 | 1.00 | Bucklin | 0.31 | 0.46 | 1.00 | 1.00 | 1.00 | 1.00 | n/a |
| 6 | 600 | 1.00 | Coombs | 0.33 | 0.44 | 0.53 | 0.65 | 1.00 | 1.00 | n/a |
| 6 | 600 | 1.00 | Plurality With Runoff | 0.57 | 0.69 | 0.81 | 0.86 | 1.00 | 1.00 | n/a |
| 6 | 600 | 1.00 | Schulze | 0.41 | 0.55 | 0.66 | 0.79 | 1.00 | 1.00 | n/a |
| 6 | 2,000 | 0.70 | Bucklin | 1.00 | 1.00 | 1.00 | 1.00 | 1.00 | 1.00 | n/a |
| 6 | 2,000 | 0.70 | Coombs | 1.00 | 1.00 | 1.00 | 1.00 | 1.00 | 1.00 | n/a |
| 6 | 2,000 | 0.70 | Plurality With Runoff | 1.00 | 1.00 | 1.00 | 1.00 | 1.00 | 1.00 | n/a |
| 6 | 2,000 | 0.70 | Schulze | 1.00 | 1.00 | 1.00 | 1.00 | 1.00 | 1.00 | n/a |
| 6 | 2,000 | 0.80 | Bucklin | 1.00 | 1.00 | 1.00 | 1.00 | 1.00 | 1.00 | n/a |
| 6 | 2,000 | 0.80 | Coombs | 1.00 | 1.00 | 1.00 | 1.00 | 1.00 | 1.00 | n/a |
| 6 | 2,000 | 0.80 | Plurality With Runoff | 1.00 | 1.00 | 1.00 | 1.00 | 1.00 | 1.00 | n/a |
| 6 | 2,000 | 0.80 | Schulze | 1.00 | 1.00 | 1.00 | 1.00 | 1.00 | 1.00 | n/a |
| 6 | 2,000 | 0.90 | Bucklin | 0.93 | 0.95 | 1.00 | 1.00 | 1.00 | 1.00 | n/a |
| 6 | 2,000 | 0.90 | Coombs | 0.94 | 0.97 | 0.99 | 0.99 | 1.00 | 1.00 | n/a |
| 6 | 2,000 | 0.90 | Plurality With Runoff | 0.94 | 0.98 | 0.99 | 0.99 | 1.00 | 1.00 | n/a |
| 6 | 2,000 | 0.90 | Schulze | 0.94 | 0.97 | 0.99 | 0.99 | 1.00 | 1.00 | n/a |
| 6 | 2,000 | 1.00 | Bucklin | 0.34 | 0.51 | 1.00 | 1.00 | 1.00 | 1.00 | n/a |
| 6 | 2,000 | 1.00 | Coombs | 0.35 | 0.44 | 0.55 | 0.68 | 1.00 | 1.00 | n/a |
| 6 | 2,000 | 1.00 | Plurality With Runoff | 0.57 | 0.72 | 0.81 | 0.88 | 1.00 | 1.00 | n/a |
| 6 | 2,000 | 1.00 | Schulze | 0.43 | 0.58 | 0.70 | 0.82 | 1.00 | 1.00 | n/a |
| 7 | 100 | 0.70 | Bucklin | 0.88 | 0.95 | 1.00 | 1.00 | 1.00 | 1.00 | 1.00 |
| 7 | 100 | 0.70 | Coombs | 0.90 | 0.94 | 0.98 | 0.99 | 1.00 | 1.00 | 1.00 |
| 7 | 100 | 0.70 | Plurality With Runoff | 0.90 | 0.95 | 0.98 | 0.99 | 1.00 | 1.00 | 1.00 |
| 7 | 100 | 0.70 | Schulze | 0.90 | 0.94 | 0.98 | 0.99 | 1.00 | 1.00 | 1.00 |
| 7 | 100 | 0.80 | Bucklin | 0.67 | 0.79 | 1.00 | 1.00 | 1.00 | 1.00 | 1.00 |
| 7 | 100 | 0.80 | Coombs | 0.74 | 0.86 | 0.90 | 0.92 | 0.95 | 1.00 | 1.00 |
| 7 | 100 | 0.80 | Plurality With Runoff | 0.78 | 0.89 | 0.92 | 0.94 | 0.97 | 1.00 | 1.00 |
| 7 | 100 | 0.80 | Schulze | 0.75 | 0.87 | 0.90 | 0.92 | 0.96 | 1.00 | 1.00 |
| 7 | 100 | 0.90 | Bucklin | 0.43 | 0.60 | 0.91 | 1.00 | 1.00 | 1.00 | 1.00 |
| 7 | 100 | 0.90 | Coombs | 0.47 | 0.60 | 0.68 | 0.75 | 0.83 | 1.00 | 1.00 |
| 7 | 100 | 0.90 | Plurality With Runoff | 0.57 | 0.72 | 0.81 | 0.85 | 0.91 | 1.00 | 1.00 |
| 7 | 100 | 0.90 | Schulze | 0.50 | 0.65 | 0.75 | 0.81 | 0.88 | 1.00 | 1.00 |
| 7 | 100 | 1.00 | Bucklin | 0.25 | 0.37 | 0.58 | 1.00 | 1.00 | 1.00 | 1.00 |
| 7 | 100 | 1.00 | Coombs | 0.25 | 0.34 | 0.42 | 0.50 | 0.62 | 1.00 | 1.00 |
| 7 | 100 | 1.00 | Plurality With Runoff | 0.49 | 0.61 | 0.71 | 0.79 | 0.86 | 1.00 | 1.00 |
| 7 | 100 | 1.00 | Schulze | 0.34 | 0.47 | 0.57 | 0.67 | 0.78 | 1.00 | 1.00 |
| 7 | 200 | 0.70 | Bucklin | 0.95 | 0.99 | 1.00 | 1.00 | 1.00 | 1.00 | 1.00 |
| 7 | 200 | 0.70 | Coombs | 0.96 | 0.98 | 0.99 | 0.99 | 1.00 | 1.00 | 1.00 |
| 7 | 200 | 0.70 | Plurality With Runoff | 0.96 | 0.98 | 0.99 | 0.99 | 1.00 | 1.00 | 1.00 |

| 7 | 200 | 0.70 | Schulze | 0.96 | 0.98 | 0.99 | 0.99 | 1.00 | 1.00 | 1.00 |
| 7 | 200 | 0.80 | Bucklin | 0.82 | 0.87 | 1.00 | 1.00 | 1.00 | 1.00 | 1.00 |
| 7 | 200 | 0.80 | Coombs | 0.85 | 0.92 | 0.95 | 0.96 | 0.98 | 1.00 | 1.00 |
| 7 | 200 | 0.80 | Plurality With Runoff | 0.86 | 0.92 | 0.96 | 0.97 | 0.98 | 1.00 | 1.00 |
| 7 | 200 | 0.80 | Schulze | 0.86 | 0.92 | 0.96 | 0.96 | 0.98 | 1.00 | 1.00 |
| 7 | 200 | 0.90 | Bucklin | 0.56 | 0.72 | 0.93 | 1.00 | 1.00 | 1.00 | 1.00 |
| 7 | 200 | 0.90 | Coombs | 0.58 | 0.71 | 0.79 | 0.84 | 0.89 | 1.00 | 1.00 |
| 7 | 200 | 0.90 | Plurality With Runoff | 0.67 | 0.79 | 0.86 | 0.91 | 0.94 | 1.00 | 1.00 |
| 7 | 200 | 0.90 | Schulze | 0.60 | 0.75 | 0.82 | 0.88 | 0.92 | 1.00 | 1.00 |
| 7 | 200 | 1.00 | Bucklin | 0.25 | 0.37 | 0.50 | 1.00 | 1.00 | 1.00 | 1.00 |
| 7 | 200 | 1.00 | Coombs | 0.28 | 0.39 | 0.45 | 0.53 | 0.64 | 1.00 | 1.00 |
| 7 | 200 | 1.00 | Plurality With Runoff | 0.54 | 0.67 | 0.75 | 0.81 | 0.89 | 1.00 | 1.00 |
| 7 | 200 | 1.00 | Schulze | 0.35 | 0.49 | 0.58 | 0.68 | 0.80 | 1.00 | 1.00 |
| 7 | 300 | 0.70 | Bucklin | 0.98 | 1.00 | 1.00 | 1.00 | 1.00 | 1.00 | 1.00 |
| 7 | 300 | 0.70 | Coombs | 0.99 | 1.00 | 1.00 | 1.00 | 1.00 | 1.00 | 1.00 |
| 7 | 300 | 0.70 | Plurality With Runoff | 0.99 | 1.00 | 1.00 | 1.00 | 1.00 | 1.00 | 1.00 |
| 7 | 300 | 0.70 | Schulze | 0.99 | 1.00 | 1.00 | 1.00 | 1.00 | 1.00 | 1.00 |
| 7 | 300 | 0.80 | Bucklin | 0.86 | 0.91 | 1.00 | 1.00 | 1.00 | 1.00 | 1.00 |
| 7 | 300 | 0.80 | Coombs | 0.88 | 0.94 | 0.96 | 0.98 | 0.98 | 1.00 | 1.00 |
| 7 | 300 | 0.80 | Plurality With Runoff | 0.89 | 0.94 | 0.97 | 0.98 | 0.98 | 1.00 | 1.00 |
| 7 | 300 | 0.80 | Schulze | 0.88 | 0.94 | 0.96 | 0.98 | 0.98 | 1.00 | 1.00 |
| 7 | 300 | 0.90 | Bucklin | 0.61 | 0.74 | 0.93 | 1.00 | 1.00 | 1.00 | 1.00 |
| 7 | 300 | 0.90 | Coombs | 0.65 | 0.76 | 0.83 | 0.87 | 0.91 | 1.00 | 1.00 |
| 7 | 300 | 0.90 | Plurality With Runoff | 0.70 | 0.81 | 0.88 | 0.91 | 0.95 | 1.00 | 1.00 |
| 7 | 300 | 0.90 | Schulze | 0.67 | 0.77 | 0.85 | 0.90 | 0.93 | 1.00 | 1.00 |
| 7 | 300 | 1.00 | Bucklin | 0.21 | 0.34 | 0.46 | 1.00 | 1.00 | 1.00 | 1.00 |
| 7 | 300 | 1.00 | Coombs | 0.28 | 0.38 | 0.43 | 0.53 | 0.65 | 1.00 | 1.00 |
| 7 | 300 | 1.00 | Plurality With Runoff | 0.51 | 0.67 | 0.72 | 0.80 | 0.88 | 1.00 | 1.00 |
| 7 | 300 | 1.00 | Schulze | 0.34 | 0.50 | 0.58 | 0.67 | 0.79 | 1.00 | 1.00 |
| 7 | 400 | 0.70 | Bucklin | 0.99 | 1.00 | 1.00 | 1.00 | 1.00 | 1.00 | 1.00 |
| 7 | 400 | 0.70 | Coombs | 1.00 | 1.00 | 1.00 | 1.00 | 1.00 | 1.00 | 1.00 |
| 7 | 400 | 0.70 | Plurality With Runoff | 1.00 | 1.00 | 1.00 | 1.00 | 1.00 | 1.00 | 1.00 |
| 7 | 400 | 0.70 | Schulze | 1.00 | 1.00 | 1.00 | 1.00 | 1.00 | 1.00 | 1.00 |
| 7 | 400 | 0.80 | Bucklin | 0.92 | 0.95 | 1.00 | 1.00 | 1.00 | 1.00 | 1.00 |
| 7 | 400 | 0.80 | Coombs | 0.94 | 0.97 | 0.98 | 0.99 | 0.99 | 1.00 | 1.00 |
| 7 | 400 | 0.80 | Plurality With Runoff | 0.94 | 0.97 | 0.98 | 0.99 | 0.99 | 1.00 | 1.00 |

| 7 | 400 | 0.80 | Schulze | 0.94 | 0.97 | 0.98 | 0.99 | 0.99 | 1.00 | 1.00 |
|---|-----|------|---------|------|------|------|------|------|------|------|
| 7 | 400 | 0.90 | Bucklin | 0.63 | 0.73 | 0.94 | 1.00 | 1.00 | 1.00 | 1.00 |
| 7 | 400 | 0.90 | Coombs | 0.67 | 0.79 | 0.85 | 0.89 | 0.92 | 1.00 | 1.00 |
| 7 | 400 | 0.90 | Plurality With Runoff | 0.73 | 0.83 | 0.89 | 0.92 | 0.96 | 1.00 | 1.00 |
| 7 | 400 | 0.90 | Schulze | 0.68 | 0.80 | 0.87 | 0.90 | 0.93 | 1.00 | 1.00 |
| 7 | 400 | 1.00 | Bucklin | 0.25 | 0.34 | 0.49 | 1.00 | 1.00 | 1.00 | 1.00 |
| 7 | 400 | 1.00 | Coombs | 0.30 | 0.37 | 0.47 | 0.54 | 0.65 | 1.00 | 1.00 |
| 7 | 400 | 1.00 | Plurality With Runoff | 0.51 | 0.67 | 0.77 | 0.82 | 0.88 | 1.00 | 1.00 |
| 7 | 400 | 1.00 | Schulze | 0.35 | 0.49 | 0.60 | 0.70 | 0.78 | 1.00 | 1.00 |
| 7 | 500 | 0.70 | Bucklin | 1.00 | 1.00 | 1.00 | 1.00 | 1.00 | 1.00 | 1.00 |
| 7 | 500 | 0.70 | Coombs | 1.00 | 1.00 | 1.00 | 1.00 | 1.00 | 1.00 | 1.00 |
| 7 | 500 | 0.70 | Plurality With Runoff | 1.00 | 1.00 | 1.00 | 1.00 | 1.00 | 1.00 | 1.00 |
| 7 | 500 | 0.70 | Schulze | 1.00 | 1.00 | 1.00 | 1.00 | 1.00 | 1.00 | 1.00 |
| 7 | 500 | 0.80 | Bucklin | 0.92 | 0.96 | 1.00 | 1.00 | 1.00 | 1.00 | 1.00 |
| 7 | 500 | 0.80 | Coombs | 0.95 | 0.99 | 0.99 | 1.00 | 1.00 | 1.00 | 1.00 |
| 7 | 500 | 0.80 | Plurality With Runoff | 0.95 | 0.99 | 0.99 | 1.00 | 1.00 | 1.00 | 1.00 |
| 7 | 500 | 0.80 | Schulze | 0.95 | 0.99 | 0.99 | 1.00 | 1.00 | 1.00 | 1.00 |
| 7 | 500 | 0.90 | Bucklin | 0.70 | 0.79 | 0.95 | 1.00 | 1.00 | 1.00 | 1.00 |
| 7 | 500 | 0.90 | Coombs | 0.72 | 0.84 | 0.88 | 0.91 | 0.94 | 1.00 | 1.00 |
| 7 | 500 | 0.90 | Plurality With Runoff | 0.77 | 0.88 | 0.91 | 0.94 | 0.96 | 1.00 | 1.00 |
| 7 | 500 | 0.90 | Schulze | 0.73 | 0.85 | 0.89 | 0.93 | 0.96 | 1.00 | 1.00 |
| 7 | 500 | 1.00 | Bucklin | 0.26 | 0.35 | 0.47 | 1.00 | 1.00 | 1.00 | 1.00 |
| 7 | 500 | 1.00 | Coombs | 0.31 | 0.39 | 0.51 | 0.57 | 0.65 | 1.00 | 1.00 |
| 7 | 500 | 1.00 | Plurality With Runoff | 0.53 | 0.67 | 0.77 | 0.84 | 0.90 | 1.00 | 1.00 |
| 7 | 500 | 1.00 | Schulze | 0.38 | 0.50 | 0.62 | 0.73 | 0.82 | 1.00 | 1.00 |
| 7 | 600 | 0.70 | Bucklin | 1.00 | 1.00 | 1.00 | 1.00 | 1.00 | 1.00 | 1.00 |
| 7 | 600 | 0.70 | Coombs | 1.00 | 1.00 | 1.00 | 1.00 | 1.00 | 1.00 | 1.00 |
| 7 | 600 | 0.70 | Plurality With Runoff | 1.00 | 1.00 | 1.00 | 1.00 | 1.00 | 1.00 | 1.00 |
| 7 | 600 | 0.70 | Schulze | 1.00 | 1.00 | 1.00 | 1.00 | 1.00 | 1.00 | 1.00 |
| 7 | 600 | 0.80 | Bucklin | 0.96 | 0.97 | 1.00 | 1.00 | 1.00 | 1.00 | 1.00 |
| 7 | 600 | 0.80 | Coombs | 0.98 | 0.99 | 1.00 | 1.00 | 1.00 | 1.00 | 1.00 |
| 7 | 600 | 0.80 | Plurality With Runoff | 0.98 | 0.99 | 1.00 | 1.00 | 1.00 | 1.00 | 1.00 |
| 7 | 600 | 0.80 | Schulze | 0.98 | 0.99 | 1.00 | 1.00 | 1.00 | 1.00 | 1.00 |
| 7 | 600 | 0.90 | Bucklin | 0.71 | 0.83 | 0.95 | 1.00 | 1.00 | 1.00 | 1.00 |
| 7 | 600 | 0.90 | Coombs | 0.74 | 0.85 | 0.91 | 0.93 | 0.95 | 1.00 | 1.00 |
| 7 | 600 | 0.90 | Plurality With Runoff | 0.78 | 0.87 | 0.92 | 0.95 | 0.97 | 1.00 | 1.00 |

| 7 | 600 | 0.90 | Schulze | 0.75 | 0.86 | 0.93 | 0.94 | 0.96 | 1.00 | 1.00 |
|---|---|---|---|---|---|---|---|---|---|---|
| 7 | 600 | 1.00 | Bucklin | 0.25 | 0.35 | 0.50 | 1.00 | 1.00 | 1.00 | 1.00 |
| 7 | 600 | 1.00 | Coombs | 0.31 | 0.39 | 0.48 | 0.54 | 0.65 | 1.00 | 1.00 |
| 7 | 600 | 1.00 | Plurality With Runoff | 0.53 | 0.66 | 0.74 | 0.81 | 0.89 | 1.00 | 1.00 |
| 7 | 600 | 1.00 | Schulze | 0.37 | 0.51 | 0.62 | 0.72 | 0.83 | 1.00 | 1.00 |
| 7 | 2,000 | 0.70 | Bucklin | 1.00 | 1.00 | 1.00 | 1.00 | 1.00 | 1.00 | 1.00 |
| 7 | 2,000 | 0.70 | Coombs | 1.00 | 1.00 | 1.00 | 1.00 | 1.00 | 1.00 | 1.00 |
| 7 | 2,000 | 0.70 | Plurality With Runoff | 1.00 | 1.00 | 1.00 | 1.00 | 1.00 | 1.00 | 1.00 |
| 7 | 2,000 | 0.70 | Schulze | 1.00 | 1.00 | 1.00 | 1.00 | 1.00 | 1.00 | 1.00 |
| 7 | 2,000 | 0.80 | Bucklin | 1.00 | 1.00 | 1.00 | 1.00 | 1.00 | 1.00 | 1.00 |
| 7 | 2,000 | 0.80 | Coombs | 1.00 | 1.00 | 1.00 | 1.00 | 1.00 | 1.00 | 1.00 |
| 7 | 2,000 | 0.80 | Plurality With Runoff | 1.00 | 1.00 | 1.00 | 1.00 | 1.00 | 1.00 | 1.00 |
| 7 | 2,000 | 0.80 | Schulze | 1.00 | 1.00 | 1.00 | 1.00 | 1.00 | 1.00 | 1.00 |
| 7 | 2,000 | 0.90 | Bucklin | 0.90 | 0.95 | 0.99 | 1.00 | 1.00 | 1.00 | 1.00 |
| 7 | 2,000 | 0.90 | Coombs | 0.91 | 0.96 | 0.98 | 0.99 | 1.00 | 1.00 | 1.00 |
| 7 | 2,000 | 0.90 | Plurality With Runoff | 0.91 | 0.96 | 0.98 | 0.99 | 1.00 | 1.00 | 1.00 |
| 7 | 2,000 | 0.90 | Schulze | 0.91 | 0.96 | 0.98 | 0.99 | 1.00 | 1.00 | 1.00 |
| 7 | 2,000 | 1.00 | Bucklin | 0.27 | 0.37 | 0.51 | 1.00 | 1.00 | 1.00 | 1.00 |
| 7 | 2,000 | 1.00 | Coombs | 0.30 | 0.40 | 0.47 | 0.54 | 0.65 | 1.00 | 1.00 |
| 7 | 2,000 | 1.00 | Plurality With Runoff | 0.57 | 0.70 | 0.77 | 0.85 | 0.91 | 1.00 | 1.00 |
| 7 | 2,000 | 1.00 | Schulze | 0.41 | 0.52 | 0.62 | 0.71 | 0.84 | 1.00 | 1.00 |